\documentclass[final,1p,times]{elsarticle} 

\usepackage{graphicx}
\usepackage{amssymb} 
\usepackage{amsthm} 
\usepackage{lineno}

\journal{Nuclear Physics A} 

\begin{document}

\begin{frontmatter} 

\title{Jet-Tagged Back-Scattering Photons For Quark Gluon Plasma Tomography}

\author[auth1]{Rainer J.\  Fries}
\address[auth1]{Cyclotron Institute and Department of Physics and Astronomy, Texas
  A\&M University, College Station, TX 77845, USA}
\author[auth2]{Somnath De}
\author[auth2]{Dinesh K.\ Srivastava}
\address[auth2]{Variable Energy Cyclotron Centre, 1/AF, Bidhan Nagar, Kolkata
  - 700064, India}

\begin{abstract} 
Direct photons are important probes for quark gluon plasma created in
high energy nuclear collisions. Various sources of direct photons in nuclear
collisions are known, each of them endowed with characteristic information about the
production process. However, it has been challenging to separate direct photon sources 
through measurements of single inclusive photon spectra and 
photon azimuthal asymmetry. Here we explore a method to identify photons
created from the back-scattering of high momentum quarks off quark gluon 
plasma. We show that the correlation of back-scattering photons with a trigger
jet leads to a signal that should be measurable at RHIC and LHC.
\end{abstract} 

\end{frontmatter} 



Photons and dileptons have long been considered good probes of the hot
and dense fireball created in high energy nuclear collisions. This is mostly
due to the fact that the probability for particles without color charge to
reinteract after their creation is negligible. Here we will mostly focus on 
photons though many of the arguments are equally valid for virtual photons 
and the dilepton pairs into which they decay.

Many sources of direct photons are known in high energy nuclear collisions:
This includes (i) hard initial and fragmentation photons \cite{Owens:1986mp},
which emerge from high momentum transfer scatterings of partons 
in the first instance of the collisions, either directly or as
bremsstrahlung off quark and gluon jets. These photon sources are also active
in collisions of protons.
After the initial phase there is a pre-equilibrium phase that can last up to 1
fm/$c$ in which (ii) pre-equilibrium photons can be emitted. The dynamics
during the pre-equilibrium phase is not well understood and estimates of photon
production under widely different assumptions are available 
\cite{Bass:2002pm,Basar:2012bp}.
There is ample evidence that eventually an equilibrated quark gluon plasma
(QGP) is created in collisions at energies available at the Relativistic Heavy
Ion Collider (RHIC) and the Large Hadron Collider (LHC). This leads to 
the emission of (iii) thermal photon radiation
\cite{Kapusta:1991qp,Baier:1991em,Aurenche:1998nw,Arnold:2001ms} which
has been used as a thermometer for the hot nuclear matter created in those 
collisions \cite{Adare:2008ab}. 
As the fireball cools and expands confinement of partons into hadrons sets in.
This could lead to (iv) photons from hadronization processes
\cite{ChenFries:2012}.
After hadronization the matter is still hot enough to emit
(v) thermal photons from the hadronic phase \cite{Kapusta:1991qp}.
Lastly we mention the source of photons that we will be interested in
isolating here, (vi) photons from interactions of fast partons in QGP. It has been
argued that interactions of fast quarks and gluons with QGP, that lead to energy loss
of those partons can also produce photons through Compton scattering,
annihilation and bremsstrahlung \cite{Fries:2002kt,Fries:2005zh,Zakharov:2004bi}. 
As in the purely electromagnetic sector Compton and
annihilation photons in QCD can be emitted in back-scattering kinematics, making them
a potentially high-yield source at large momenta. In these proceedings we
discuss how these back-scattering photons could be separated from other
sources and what information is encoded in them \cite{Fries:2012de}.


Elastic photon-electron back-scattering is a well studied process in quantum
electrodynamics with many technological applications. The phenomenon that
the cross section for $\gamma + e \to \gamma + e$ exhibits a sharp peak in
backward direction (i.e.\ the final photon going into the direction of the initial
electron) can be used to create intense beams of high energy photons for
experiments in material science, nuclear and particle physics  
\cite{Milburn:1962jv,Arutyuninan:1963aa}. Typically $\sim$ eV photons from an
intense laser beam are accelerated to $\sim$ GeV energies through 
back-scattering from an electron beam of large energy. In quantum
chromodynamics the mixed process $g +q \to \gamma + q$ involves the same 
diagrams at leading order and exhibits the same Compton back-scattering peak. 
A similar peak can be observed for the annihilation process $q+\bar q \to \gamma+ g$.
In \cite{Fries:2002kt} some of us have estimated that a significant yield of
high energy photons can be expected from quarks ($E_q \sim 1$ -- 100 GeV) 
scattering off thermal gluons ($E_g \sim 200$ MeV) or thermal antiquarks in
QGP.
The rate of high energy Compton back-scattering and annihilation photons from 
high momentum quarks interacting with the medium is \cite{Fries:2002kt}
\begin{equation}
\label{eq:1}
  E_\gamma \frac{dN}{d^4x d^3p_\gamma} = \frac{\alpha\alpha_s}{4\pi^2}
  \sum_{f=1}^{N_f} \left(\frac{e_f}{e}\right)^2 
  \left[ f_q(\mathbf{p}_\gamma,x) + f_{\bar q}(\mathbf{p}_\gamma,x) \right]
  T^2 \left[ \ln \frac{3E_\gamma }{\alpha_s\pi T} + C\right]
\end{equation}
where $C=-1.916$. $\alpha$ and $\alpha_s$ are the electromagnetic and strong 
coupling constant, $T$ is the local temperature at $x$, $f_q$ is the phase
space distribution of fast quarks interacting with the medium and $e_f$ is the 
electric charge of a quark with the index $f$ running over all active quark flavors. 
This rate has subsequently also been calculated for virtual photons 
\cite{Srivastava:2002ic,Turbide:2006mc}.

We note that the photon rate is proportional to $T^2 \ln 1/T$, and hence the
total yield of back-scattering photons is rather sensitive to the medium temperature.
We can also see how the time evoluation of the high momentum quark
distribution $f_q$ enters the yield linearly which leads to a power-law spectrum. Thus we
might be able to go to large transverse momentum $P_T$ to measure back-scattering
photons. 
Because back-scattering happens throughout the time a parton propagates
through QGP the rate is also sensitive to the energy loss of partons with time
through $f_q$.
It has been shown that back-scattering photons are an important contribution 
to the single inclusive direct photon spectrum at intermediate $P_T$ of a few GeV/$c$
\cite{Fries:2002kt,Fries:2005zh,Turbide:2007mi}.
However, the background of hard initial and fragmentation photons 
(dominant at larger $P_T$) and thermal photons (at lower $P_T$) is difficult
to subtract. The search for photons from jet-medium interactions in single
inclusive spectra has therefore been inconclusive. The same is true for the
typical signature that jet-medium photons would leave on the azimuthal
asymmetry $v_2$. It has been proposed that photons from jet-medium
interactions have negative $v_2$ \cite{Turbide:2005bz}, unlike hard initial
and thermal photons \cite{Chatterjee:2005de} which exhibit vanishing or positive $v_2$. 
Again the experimental situation so far is inconclusive \cite{Turbide:2007mi}. 

Since back-scattering photons are produced by fast partons from a jet
and jets prefer to be produced in back-to-back pairs we propose to use a jet
trigger to look for an unambiguous signature. Specifically we suggest to
trigger on a jet with an energy $E_\gamma$ of a few 10 GeV and to look for
direct photons in a narrow region in azimuthal angle $\phi$ on the away-side.
Only hard initial photons and fragmentation photons have a similar correlation
with a jet on the opposite side. These will be the sources that create the 
background for the measurement. All thermal and pre-equilibrium sources do not
have a correlation with an away-side jet and can be neglected from the outset.
Next we notice that fragmentation photons are concentrated at  low $z$ ($\lesssim 0.3$)
where $z=E_\gamma/E_{\mathrm{parent\,\, jet}}$
\cite{Owens:1986mp,Bourhis:2000gs} .  Thus we can suppress the background from
fragmentation (and induced bremsstrahlung) by choosing  values of $E_\gamma$
close to the trigger jet energy. We recall that at leading order (LO)
kinematics the transverse momenta of the trigger jet and the photon
or parton on the other side are perfectly balanced. 
Energy loss before the conversion will shift the energy of 
back-scattering photons away from the trigger energy and thus  to lower
momentum compared to the background of hard initial
photons. Beyond leading order the back-to-back correlation of
both the back-scattering signal and the background (hard and
fragmentation) photons are somewhat softened.


We are now going to present a numerical feasibility study. We use the 
JETPHOX code \cite{Catani:2002ny,Aurenche:2006vj} to calculate jet-photon 
and jet-hadron yields at LO and next-to-leading order (NLO) accuracies.
Trigger jets were fixed around midrapidity in rather narrow windows in
transverse energy. Then the
photon spectrum in a sector of $\pm 15^\circ$  around the away-side is
considered. We use the PPM code \cite{Rodriguez:2010di,Fries:2010jd} 
to propagate leading jet partons of the away-side jet through a 
longitudinally boost-invariant fireball parameterization whose entropy
$dS/dy$ has been tuned to measured RHIC and LHC multiplicities.
PPM calculates the energy loss of partons and the back-scattering process
to create photons. The energy loss is tuned to describe measured single
inclusive hadron spectra.

\begin{figure}[htbp]  
\begin{center}
  \includegraphics[width=0.45\textwidth]{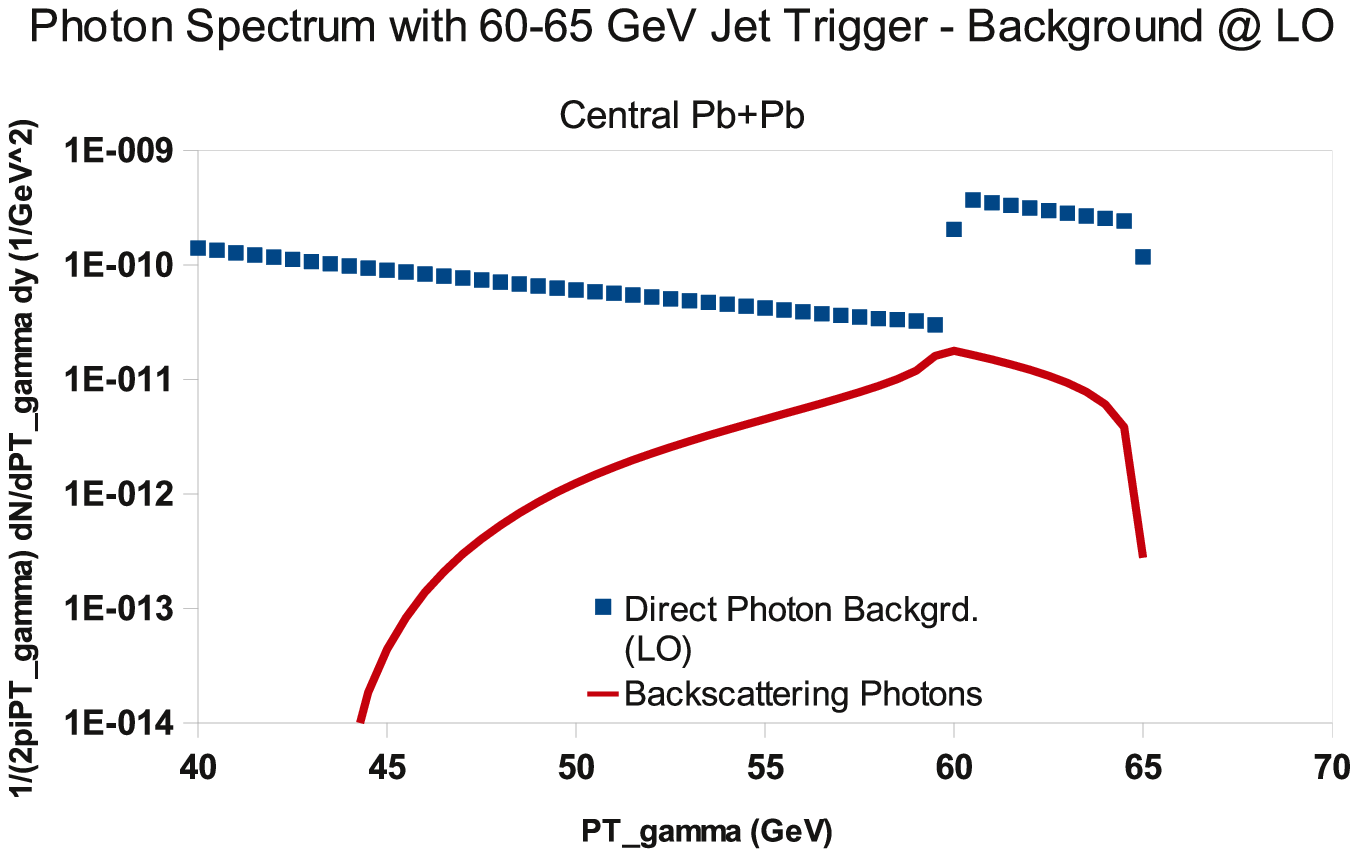}
  \includegraphics[width=0.45\textwidth]{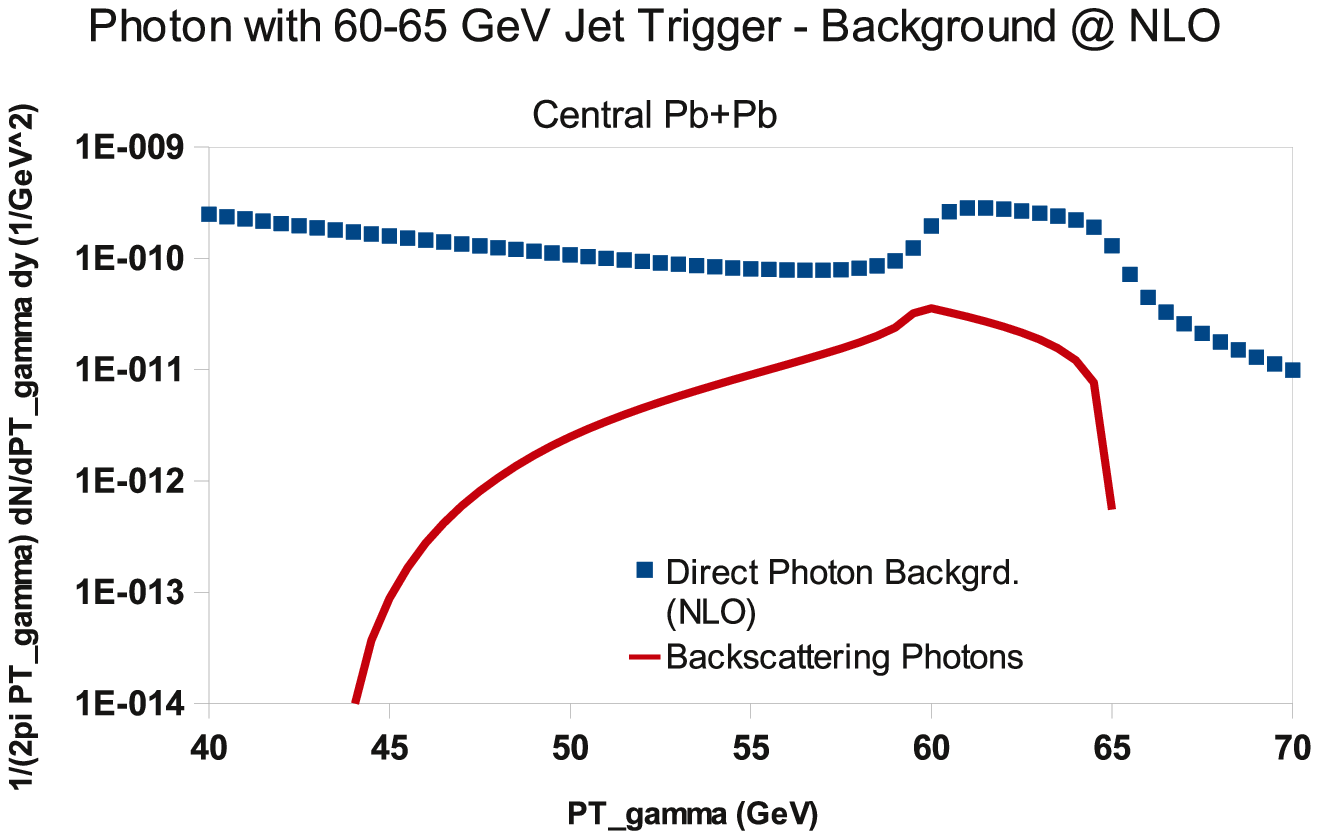}
\end{center}
  \caption{
    Left panel: photon spectra associated with trigger jets from 60-65
    GeV in central Pb+Pb collisions at LHC at $\sqrt{s_{NN}}=2.76 $ TeV at LO
    accuracy. The yield of back-scattering photons (red solid line)
    is compared to the background of hard initial photons and fragmentation 
    photons (blue squares). Right
   panel: same with background calculated at NLO accuracy and signal scaled
   with a $K$-factor of 2.}
\label{fig:1}
\end{figure}

\begin{figure}[htbp]  
\begin{center}
  \includegraphics[width=0.45\textwidth]{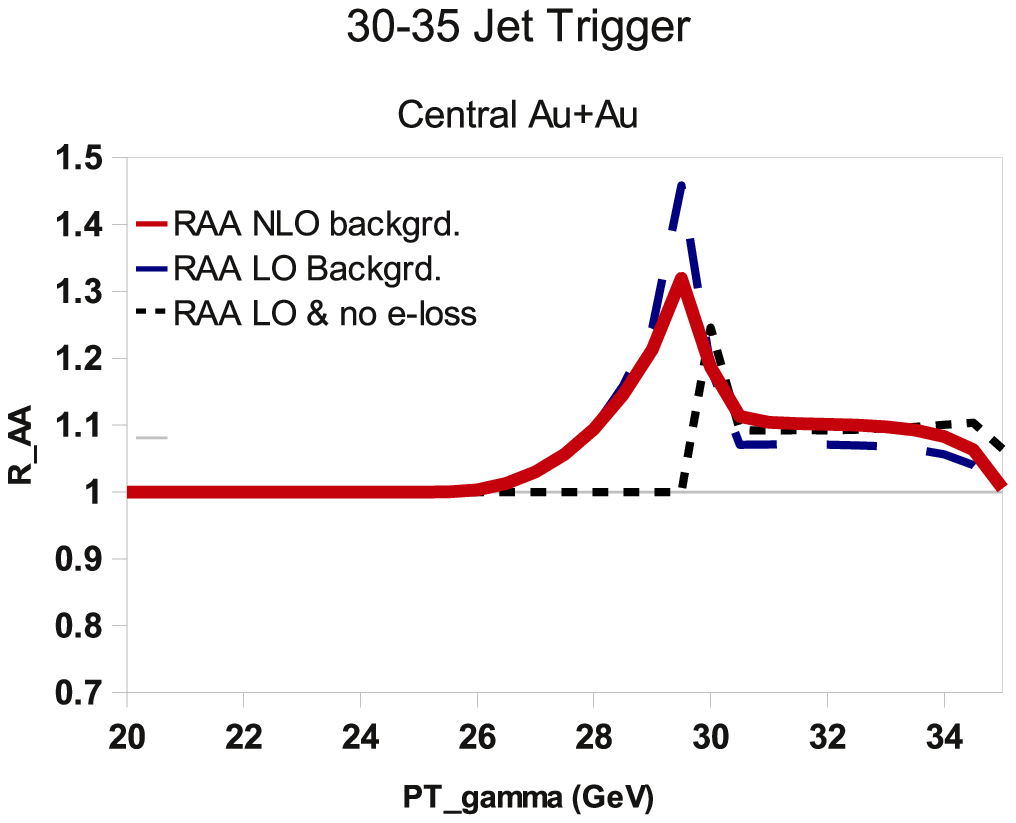}
  \includegraphics[width=0.45\textwidth]{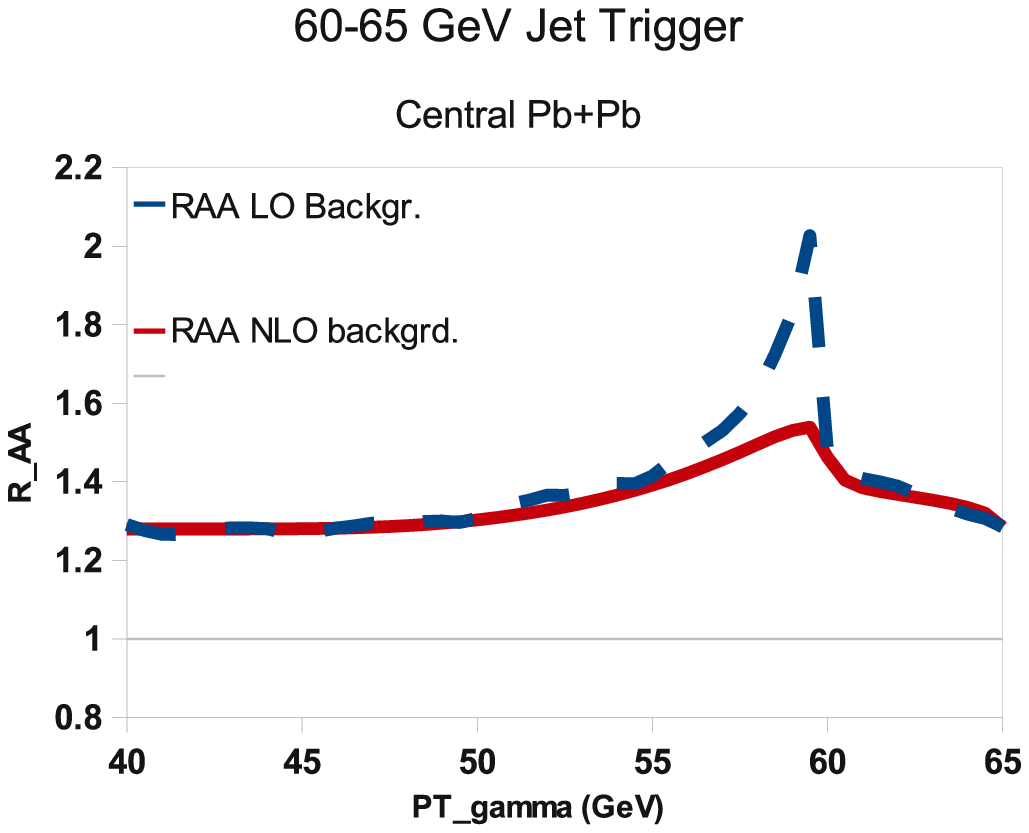}
\end{center}
  \caption{Left panel: 
    The nuclear modification factor $R_{AA}$ at LO accuracy for central Au+Au
    collisions at $\sqrt{s_{NN}}=200 $ GeV with trigger jet energies between
    30 and 35 GeV with (blue dashed line) and without (black dotted line)
    taking energy loss of leading partons into
    account. The NLO result with energy loss is also shown (red solid line).
   Right panel: LO and NLO results for the LHC case from Fig.\ \ref{fig:1}.}
\label{fig:2}
\end{figure}

Fig.\ \ref{fig:1} shows typical results for LHC energies for a trigger jet window of
$E_T=60$-$65$ GeV/$c$, once in LO kinematics and for comparison also with
the background calculated at NLO accuracy and the signal scaled with an
appropriate $K$-factor of 2. We notice the strong correlation of the signal
with the initial hard photon peak in the trigger window. However energy loss
leads to a tail of the signal extending to lower momenta. In Fig.\ \ref{fig:2} we
show the nuclear modification factor $R_{AA}$ which we approximate by the
ratio of signal (back-scattering) over the sum of signal and background (initial
hard and fragmentation). Both the LHC case from Fig.\ \ref{fig:1} and a case for
RHIC energies with a trigger jet energy $E_T=30$-$35$ GeV/$c$ are shown.
$R_{AA}$ is the most promising quantity to look for a
back-scattering signature since we see a characteristic peak just below
the trigger jet window. NLO kinematics washes out the signal but keeps it
recognizable.


In summary, we have shown that the use of trigger jets facilitates the separation of
back-scattering photons from other direct photons sources. For a trigger jet
of a few 10 GeV energy direct photons on its away-side come mainly from
initial hard processes and fragmentation of an away-side jet. However, 
back-scattering photons from the leading quark of an away-side jet lead
to a characteristic enhancement of the nuclear modification factor for direct
photons a few GeV below the trigger energy. The height of the back-scattering
peak reflects the the temperature of the medium in which the back-scattering
occurs while the shift in energy compared to the trigger jet energy is
sensitive energy loss of the quark before back-scattering.

This work was supported by NSF CAREER Award PHY-0847538 and by the JET
Collaboration and DOE grant DE-FG02-10ER41682.


\end{document}